\documentclass[twocolumn,showpacs,amsfonts]{revtex4}
\usepackage{latexsym}
\usepackage{float}
\usepackage{amssymb}
\usepackage{graphicx}
\usepackage{subfigure}
\usepackage{epsfig}
\usepackage{textcomp}
\usepackage{hyperref}
\usepackage{tabularx}

\begin{document}

\title{Aging and Crossovers in Phase-Separating Fluid Mixtures}

\author{Shaista Ahmad$^{1,2}$, Federico Corberi$^{3}$, 
Subir K. Das$^{1,*}$, Eugenio Lippiello$^{4}$, 
Sanjay Puri$^{2}$ and Marco Zannetti$^{3}$}

\affiliation{$^1$Theoretical Sciences Unit, Jawaharlal Nehru Centre for
Advanced Scientific Research, Jakkur P.O., Bangalore 560064, India \\
$^2$School of Physical Sciences, Jawaharlal Nehru University, New Delhi 110067, India \\
$^{3}$Dipartimento di Matematica e Informatica and CNISM, Unit\`a di Salerno,
Universit\`a di Salerno, via Ponte don Melillo, 84084 Fisciano (SA), Italy \\
$^{4}$Dipartimento di Scienze Ambientali, Seconda Universit\`a di Napoli,
Via Vivaldi, Caserta, Italy}

\date{\today}

\begin{abstract}
We use state-of-the-art molecular dynamics simulations to study hydrodynamic effects on aging during kinetics of phase separation in a fluid mixture. The domain growth law shows a crossover from a 
{\it diffusive} regime to a {\it viscous hydrodynamic} regime. 
There is a corresponding crossover in the autocorrelation function from a 
power-law behavior to an exponential decay. While the former is
consistent with theories for diffusive domain growth, the latter
results as a  
consequence of faster advective transport in fluids for which an analytical justification
has been provided.
\end{abstract}

\maketitle

\section{Introduction}

There has been intense recent interest in understanding the coarsening dynamics of phase-separating mixtures \cite{Dan,Ahmad,Majumder,Khanna,Lippiello,Hore,Bucior,Sicilia,Blond,Mitchel,Das1,
Castellano,Salernogroup}.
For fluid systems, attention has focused on single-time quantities, e.g., 
the correlation function and structure factor, or the growth law of the domain scale $\ell(t)$ \cite{Ahmad,Mitchel,Puri1,Bray}. An equally important and deeply related 
aspect of coarsening systems is the nature
of the aging properties \cite{Cugliandolo} which are encoded in two-time quantities. This issue has been 
intensively studied in the context of disordered systems and structural glasses, 
in phase-separating systems without hydrodynamic modes, such as solid mixtures, but not in segregating fluids. 
In part, this is due to the 
scarcity of reliable numerical results for these systems. In this letter, we present the first molecular dynamics (MD) 
results for aging in coarsening binary fluid. Before discussing our model and results, 
it is useful to give a brief overview of the main concepts.

Following a quench inside the miscibility gap, a homogeneous binary mixture (A+B) separates into A-rich and B-rich
phases. The evolution of the system from the randomly-mixed phase to the segregated state is a complex process. Domains rich in A- and B-particles form and grow nonlinearly with time \cite{Puri1,Bray,Binder1,Jones}. This coarsening is a self-similar process which is reflected in the scaling behavior of various physical quantities, as the two-point equal-time 
correlation function 
$G(r,t)=\langle \phi (\vec r_1,t)\phi(\vec r_2,t)\rangle-\langle \phi (\vec r_1,t)\rangle \langle \phi(\vec r_2,t)\rangle$ 
(where the order parameter $\phi$
is the difference between the A and B concentrations and
$r=|\vec r_1-\vec r_2|$).
This quantity can be expressed in the scaling form
$G(r,t) = \tilde {G}(r/\ell)$,
$\tilde G(z)$ being a time-independent master function. Typically, $\ell(t)$ grows in a power-law manner as
$\ell(t)\sim t^\alpha$,
where the growth exponent $\alpha$ depends upon the transport mechanism. For diffusive domain growth \cite{LS} $\alpha=1/3$, which is referred to as the Lifshitz-Slyozov law. This behavior characterizes segregating solid mixtures. However in fluids, hydrodynamic transport mechanisms, important at large length scales, lead to much
faster growth. In spatial dimension $d=3$, for a percolating domain morphology, advective transport results in three growth regimes \cite{Siggia}:
\begin{equation}\label{GL}
\alpha = \left\{
\begin{array}{l l}
1/3, & \quad \ell(t) \ll \ell_{dv}={(D\eta)}^{1/2},\\
1, & \quad \ell_{dv} \ll \ell(t) \ll \ell_{in}=\eta^2/(\rho\gamma), \\
2/3, & \quad \ell(t) \gg \ell_{in}.\\ 
\end{array}
\right .
\end{equation}
In Eq.~(\ref{GL}), $D$ is the diffusion constant, $\eta$ the viscosity, $\rho$ is the density, $\gamma$ the surface 
tension, $\ell_{dv}$ is the crossover length from diffusive to viscous regime, and
$\ell_{in}$ the one from viscous to inertial hydrodynamic regime \cite{Bray}.

For diffusive phase-separation,
also the order-parameter {\it auto-correlation function} 
$C(t,t_w)=\langle \phi(\vec r,t)\phi (\vec r,t_w)\rangle-\langle \phi(\vec r,t)\rangle \langle \phi (\vec r,t_w)\rangle$
(where $t$ and the {\it waiting time} $t_w<t$ are two generic instants)
takes a scaling form
$C(t,t_w)=\tilde C (\ell/\ell_w)$
where hereafter $\ell=\ell(t)$ and $\ell_w=\ell(t_w)$.
The scaling function behaves as $\tilde C(x)=x^{-\lambda}$
for large $x=\ell/\ell_w$, $\lambda$ being the Fisher-Huse (FH) exponent 
\cite{Huse} whose value in $d=3$, according to the results in Ref. \cite{YRD}, is in the range $[1.5,1.6]$.
Despite this good understanding 
of the aging properties in diffusive systems, the behavior in segregating
fluids remains largely unknown. In this letter we start filling the gap by studying 
the properties of the autocorrelation function in a range of times encompassing 
the first two regimes of Eq. (\ref{GL}) in a model of binary fluid.
In doing that we uncover an interesting structure with
a crossover which corresponds to the switch from the early diffusive to the
intermediate hydrodynamic regime in Eq.~(\ref{GL}) for the growth law. 
Specifically, we find the two parameter scaling form
\begin{equation}\label{autosc}
C(t,t_w,\ell_{dv})=\tilde C(x,y),
\end{equation}
where $x=\ell/\ell_w$, $y=\ell_{dv}/\ell_w$,
and $\tilde C(x,y)$ obeys the limiting behaviors:
\begin{equation}\label{limitsc}
\tilde C(x,y) = \left\{
\begin{array}{l l}
 x^{-\lambda}\,\, (\mbox{with}\, \lambda \simeq 1.5),\,\,\,\,\,\,y\gg 1; \\
 e^{-(x-1)/\tau}, \quad \quad \quad \quad \quad  y\ll 1. \\
\end{array}
\right .
\end{equation}
Thus, in the diffusive regime ($\ell_w\ll \ell_{dv}$ or $y \gg 1$), we obtain a power-law decay of the
scaling function as for spin systems. To the best of our knowledge, 
this has not been observed in previous studies of fluid phase separation. Furthermore, and more
importantly, in the hydrodynamic regime ($\ell_w \gg \ell_{dv}$ or $y \ll 1$), we find a crossover to a
novel exponential. The latter finding is the central result of this letter. Here
$y\gg 1$ and $y\ll 1$ should be read as nearly pure diffusive and pure hydrodynamic regimes, respectively.

The rest of the paper is organized as follows. We describe the model and methodology in Section II. Results
are presented in Section III. Finally, Section IV concludes the paper with a brief summary.

\section{Model and Method} 

We consider a periodic box of linear dimension $L$ containing A and B particles 
of equal mass $m$. Particles $i$ and $j$, at distance $r$ apart, interact via the potential
\begin{equation}\label{LJfull}
\left \{
\begin{array}{ll}
u(r) = U(r)-U(r_{c})-
(r-r_{c})\left(\frac{dU}{dr}\right)_{r=r_c};~r<r_c, \\
u(r)=0;~r>r_c .
\end{array}
\right .
\end{equation} 
Here, the Lennard-Jones pair potential $U(r)$ has the form
$U(r)=4\varepsilon_{\alpha\beta}[(\sigma/r)^{12} -
(\sigma/r)^{6}]$, where
$\sigma$ is the interaction diameter, and $\varepsilon_{\alpha\beta}$ is the
interaction strength between particles of species $\alpha$ and $\beta$ [$\alpha,\beta\in (A,B)$]. 
The potential in Eq.~(\ref{LJfull}) is cut-off at $r=r_c$, to ensure faster computation \cite{allen,frenkel}. 
For convenience, we set $r_c=2.5\sigma$. The third term
on the right-hand-side of Eq.~(\ref{LJfull}), after a shifting 
of the potential by its value at $r=r_c$ (see the second term), was introduced to make
both the potential and force continuous at $r=r_c$. The overall density $\rho$
is set to unity so that the fluid is incompressible. For the choice
$\varepsilon_{AA}=\varepsilon_{BB}=2\varepsilon_{AB}=\varepsilon$,
we have an Ising-like fully symmetric model. The phase behavior and equilibrium properties of this system, that exhibits a liquid-liquid phase transition at a critical temperature $k_BT_c\simeq 1.423\varepsilon$, are well studied \cite{Das2}. For convenience, we set $\varepsilon$, $k_B$, $m$ and $\sigma$ to unity below. Our choice of a high density ensures that the gas-liquid transition is well separated from the liquid-liquid one.

We employ this model to study the kinetics of phase separation via MD simulations \cite{allen,frenkel} by quenching homogeneous critical (50\% A and 50\% B particles) configurations, prepared at a very high temperature $(T=10)$, 
to temperatures below $T_c$. We have used the Verlet velocity algorithm with integration time step 
$\Delta t = 0.01\tau$, where the time unit $t_0=(m\sigma^2/\varepsilon)^{1/2}=1$. 
The temperature was controlled via application of a Nos\'{e}-Hoover thermostat  \cite{frenkel}, 
that is well-known to preserve hydrodynamics.

While unambiguous confirmation of expected hydrodynamic effects, via MD simulation, in binary fluid phase separation
was done only recently, by using this model \cite{Ahmad, new}, signature of such fast
hydrodynamic mechanism was observed in a number of earlier MD simulations \cite{lara, kab, thak}.

For analysis of our results, we mapped the continuum fluid configurations 
onto a simple cubic lattice of size $L^3$ \cite{Ahmad}. There each site $i$ is occupied
by a particle and we assign 
a spin value $S_i= +1$ if it is an A-particle
and $S_i=-1$  elsewhere.
This mapping ensures that the pattern dynamics can be studied in a manner analogous 
to the spin-$1/2$ Ising model. The two-point equal-time correlation function was calculated as 
$G(r,t)=\langle S_i(t)S_j(t) \rangle-\langle S_i(t) \rangle \langle S_j(t) \rangle$,
with $r=|i-j|$ 
and the angular brackets denote an averaging over independent runs. For a conserved order-parameter system, $G(r,t)$ exhibits damped oscillations around zero. The average domain size $\ell(t)$ is obtained from the first zero crossing of $G(r,t)$. The auto-correlation function was calculated as
$C(t,t_w)=\langle S_i(t)S_i(t_w)\rangle - \langle S_i(t) \rangle \langle S_i(t_w) \rangle$.
All results presented subsequently for $T=1.1=0.77\, T_c$ correspond to $L=64$, whereas those for $T=1.25=0.88\, T_c$ were obtained for $L=48$. For statistical quantities, we average over 10 independent runs.

\begin{figure}[htb]
\centering
\includegraphics*[width=0.375\textwidth]{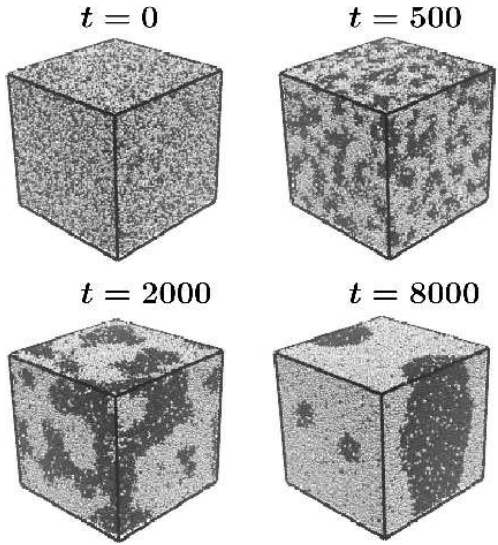}
\caption{\label{fig1} Evolution snapshots obtained after a quench  to a temperature $T=1.1$. The A-particles are marked black, B-particles are gray.}
\end{figure}

\section{Results} 

Figure~\ref{fig1} shows snapshots from the evolution of the binary 
fluid. 
The snapshot at $t=0$ corresponds to the homogeneously mixed phase immediately after the quench
to the temperature $T=1.1$. The other snapshots show the growth of bicontinuous A-rich and B-rich domains -- this continues until the system reaches equilibrium.

\begin{figure}[htb]
\centering
\includegraphics*[width=0.375\textwidth]{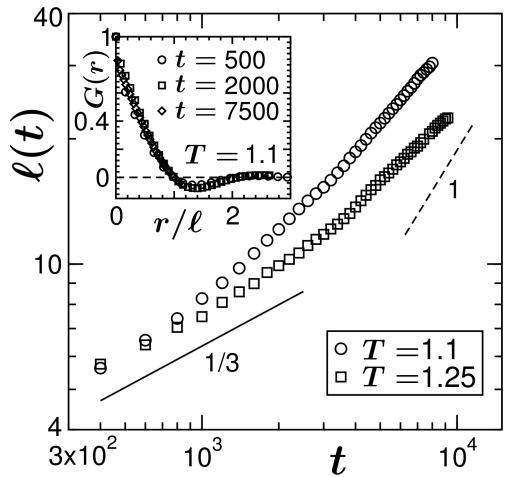}
\caption{\label{fig2}
Growth $\ell(t)$ as a function of time for temperatures $T=1.1$ and $T=1.25$, on a log-log scale. The solid and dashed lines correspond to diffusive ($\alpha = 1/3$) and viscous hydrodynamic ($\alpha = 1$) growth laws, respectively. Inset: Scaling plot of
$G(r,t)$ from three different times for $T=1.1$.}
\end{figure}

To obtain a quantitative understanding of the evolution, in Fig.~\ref{fig2} we plot $\ell (t)$ for two different temperatures \cite{Ahmad}. 
The inset shows the collapse of data for $C(r,t)$ from different times 
upon scaling the abscissa by $\ell (t)$. 
The behavior of $\ell (t)$ is consistent with the expected diffusive behavior (with $\alpha=1/3$) at early times,
with a crossover to a faster growth later on.  
As discussed in~\cite{new}, the second regime is consistent with linear hydrodynamic growth at later times. 
Indeed
the slightly off-parallel nature (on a log-log plot) of the late-time data from the linear behavior is due 
to non-zero off-sets at the crossovers 
that can be conveniently subtracted off to obtain a genuine linear behavior \cite{new}.
We observe that the crossover between the two growth regimes is seen to occur earlier
for lower temperatures, since diffusion is enhanced by thermal fluctuations. 

\begin{figure}[htb]
\centering
\includegraphics*[width=0.375\textwidth]{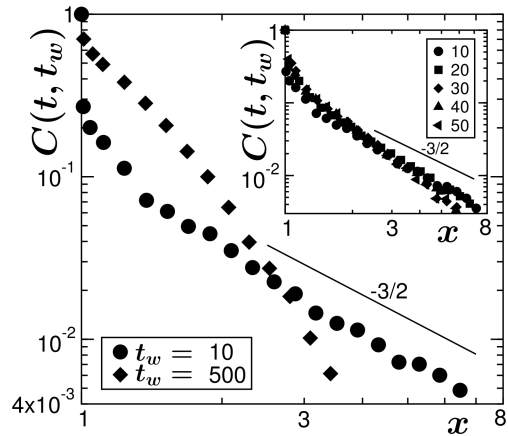}
\caption{\label{fig3} Plot of $C(t,t_w)$ vs. $x=\ell(t)/\ell(t_w)$ for $T=1.1$ and different values of $t_w$ 
(Main: $t_w= 10,500$, Inset: $t_w=10,20,30,40,50$), on a log-log scale. The solid lines denote the power-law behavior 
$x^{-3/2}$.}
\end{figure}

\begin{figure}[htb]
\centering
\includegraphics*[width=0.375\textwidth]{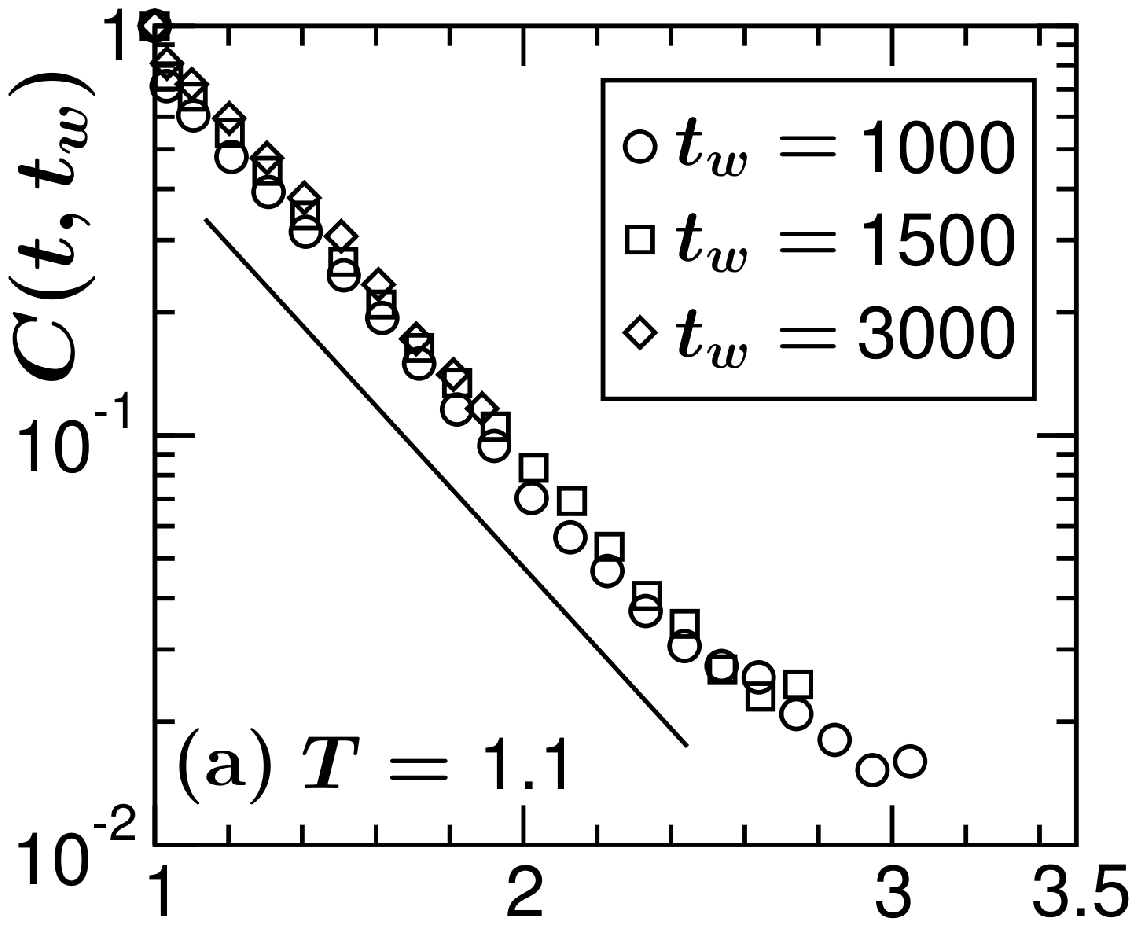}
\includegraphics*[width=0.375\textwidth]{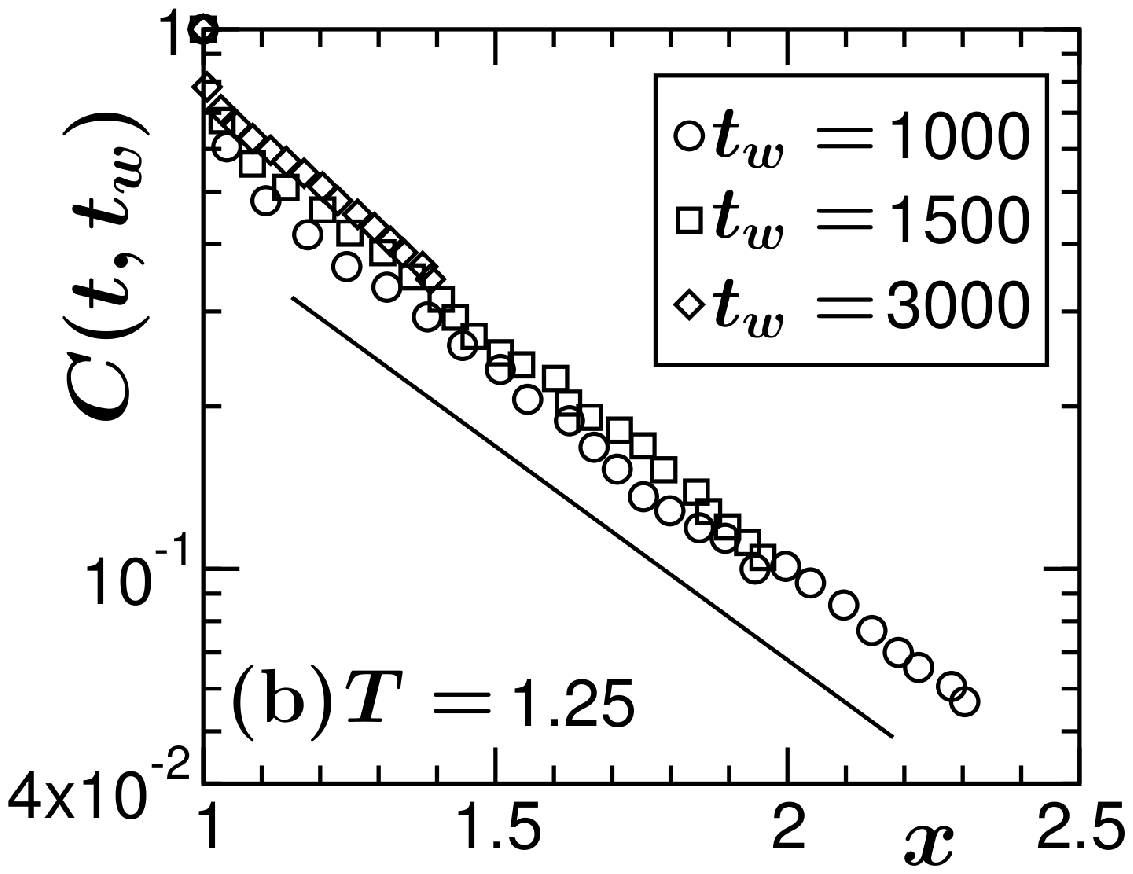}
\caption{\label{fig4} 
(a) Same as Fig.\ref{fig3} but on a semi-log plot and for $t_w=1000,1500,3000$. (b) As in (a) but at $T=1.25$.}
\end{figure}

As stated earlier, our primary objective is to study aging phenomena, essentially the
behavior of $C(t,t_w)$. In Fig.~\ref{fig3}, we plot $C(t,t_w)$ as a function of 
$x=\ell (t) /\ell (t_w)$ for different $t_w$. 
It is observed that, for the smaller values of $t_w$, presented in the inset, 
there is a reasonable collapse of data, not perfect though, and the master-curve obeys the form 
$\tilde C(x) \sim x^{-\lambda}$, with 
$\lambda \simeq 1.5$.  This is the expected behavior for phase-separation in the diffusive stage. 
We must add here the reason for the imperfect collapse of data which will also explain the logic behind the choice of
a narrow range of $t_w$ to demonstrate this power-law scaling behavior. The height at the beginning of the
power-law regime changes with $t_w$, the increase of which takes the domain order-parameter closer to the
equilibrium value. Further, with the increase of $t_w$, faster decay due to hydrodynamics (see below) shows up earlier in $x$.
Due to this latter fact the data sets as a whole give the impression of an exponent steeper than the
expected value. However, from the results for $t_w=10$ in the main frame, consistencey with the FH exponent should
be appreciated.

On the other hand, for larger values of $t_w$, e.g. $t_w=500$ in the main frame, a crossover from the FH behavior is observed, where, in the
post crossover regime
the decay of $C(t,t_w)$ is much faster than a power law.
In order to check for a possible exponential decay
at large $t_w$, a semi-log plot is presented in Fig.~\ref{fig4}(a). 
In this figure, the values of $t_w$ were judicially chosen such that for time beyond this, diffusive mechanism
is practically negligible compared to the influence of hydrodynamics. The smallest value of $t_w$ in this figure,
viz., $t_w=1000$, was chosen being motivated by a finite-size scaling study \cite{new} that showed a pure viscous 
hydrodynamic behavior for time beyond this.
Interestingly, in this post-crossover regime a scaling form is recovered, as proven by the excellent
data collapse. Moreover, the behavior of the scaling function $\tilde C(x)$ is very nicely consistent with an exponential decay. 
A similar behavior, with scaling and 
exponential decay of $\tilde C(x)$ 
is also observed for $T=1.25$ (Fig.~\ref{fig4}(b)), showing that the whole
pattern is rather general. 
Here we note that the rate of decay, defined by $\tau$, in this hydrodynamic regime should depend upon the interfacial tension
and the advection field (influenced by transport properties like shear viscosity). This will be justified towards
the end of this paper.
As is clear from parts (a) and (b)
of Fig.\ref{fig4}, this decay rate is a temperature dependent quantity. 

The deviation from a linear look, observed for $T=1.1$ at very large values of $x$, 
occur in a range of time where finite-size effects have been observed \cite{new}.
Indeed, as it can be recognized in Fig. \ref{fig1},
for such times the size of domains is comparable to the system size,
the liquid is close to equilibrium and thus the configurations change very slowly. For this reason, we are unable 
to access the {\it inertial hydrodynamic} regime [$\alpha=2/3$ in Eq.~(\ref{GL})]
which would require significantly larger system sizes 
and is beyond our 
present computational resources. 

We attribute this exponential decay of the autocorrelation function to a fast advective field in
the hydrodynamic regime which is explained below. The continuum dynamical equation [Cahn-Hilliard], used
to study kinetics of diffusive phase separation in binary alloys, can be modified to \cite{Bray}
\begin{equation}\label{explain1}
\frac{\partial \phi}{\partial t}+{\vec v}\cdot{\vec \nabla}{\phi} =D\nabla^2\mu,
\end{equation}
to investigate hydrodynamic effects in fluid phase separation. Here $\vec{v}$ is the time and space dependent
velocity field that also has dependence upon shear viscosity as well as other transport properties and
$\mu$ is the chemical potential. In the regime where hydrodynamics is the dominant mechanism,
the right hand side of Eq. (\ref{explain1}) (that takes care of diffusion) can be neglected to write
\begin{equation}\label{explain2}
\frac{\partial \phi}{\partial t}+\vec v \cdot \vec \nabla \phi=0.
\end{equation}
Assuming $\vec v$ to be a constant and representing $\phi$ in the
$k$-space, one obtains the solution
\begin{equation}\label{explain3}
\phi(\vec k,t)=\phi (\vec k,t_w)\,e^{i\vec v\cdot \vec k (t-t_w)}.
\end{equation}
Hence for the $k$-space autocorrelation
${\cal C}(\vec k,t)$ [$=\langle \phi (\vec k,t)\phi (-\vec k,t_w)\rangle$],
one has
\begin{equation}\label{explain4}
{\cal C}(\vec k,t,t_w)={\cal G}(\vec k,t_w) \,e^{i\vec v\cdot \vec k (t-t_w)},
\end{equation}
where ${\cal G}(\vec k,t_w)={\cal C}(\vec k,t_w,t_w)$ is the equal-time structure factor for
time $t_w$.
The autocorrelation function then is
\begin{eqnarray}\label{explain5}
C(t,t_w)=\int d\vec k \, {\cal C}(\vec k,t,t_w) &=& \int d\vec k \,
{\cal G}(\vec k,t_w) \,e^{i\vec v\cdot \vec k (t-t_w)}\nonumber\\
&=&
G(\vec r=\vec v (t-t_w),t_w),
\end{eqnarray}
where $G(\vec r,t_w)$ is the (real space) correlation function at time
$t_w$. Recalling the scaling property
$G(\vec r,t)=\tilde{G} \left ( \frac{r}{\ell(t)}\right )$,
one arrives at
\begin{equation}\label{explain6}
C(t,t_w)=\tilde{G} \left ( \frac{v (t-t_w)}{\ell(t_w)} \right ).
\label{final}
\end{equation}
For small values of $x$, Porod's law \cite{Bray} states that
$\tilde{G} (x)\simeq 1-ax$, where $a$ is a constant.
This is consistent with an exponential decay for small $x$.
The question then is what happens for large values of $x$. Here note that an analytical form of $\tilde{G} (x)$,
for conserved order-parameter dynamics, still remains a challenging task. However, our analysis of
the numerical data [see the inset of Fig.\ref{fig2}] is suggestive of an exponential decay of the
two-point equal time correlation function for large enough values of $x$.

\section{Conclusion} 

We have undertaken the first study of aging dynamics during phase separation in a binary Lennard-Jones fluid. The average domain size grows in a power-law fashion, with an exponent $\alpha = 1/3$ (diffusive) at early times and $\alpha = 1$ (viscous hydrodynamic) at later times. This crossover has a remarkable consequence for the scaling form of the autocorrelation function $C(t,t_w)$. In the diffusive regime, $C(t,t_w)$ shows a power-law decay with the Fisher-Huse exponent.
However, in the hydrodynamic regime, we 
observe an exponential decay. This is due to the fact that advective hydrodynamic flows 
wash out the memory very rapidly by displacing the domains. 
We believe that the results presented here will arise 
fresh experimental and theoretical interest in this problem.

\section*{Acknowledgments:} 

SKD and SA acknowledge grant number SR/S2/RJN-13/2009 of
the Department of Science and Technology, India. SKD also acknowledges 
hospitality at the University of Salerno, Italy. SA is grateful to the University 
Grants Commission for partial support and JNCASR for supporting her extended visits.\\
$^*$das@jncasr.ac.in


\begin{thebibliography}{100}

\bibitem{Dan}
D. Reith, K. Bucior, L. Yelash, P. Virnau and K. Binder, J. Phys: Cond. Mat. \textbf {24}, 115102 (2012).

\bibitem{Ahmad}
S. Ahmad, S.K. Das and S. Puri, Phys. Rev. E \textbf{82}, 040107 (2010); ibid \textbf{85}, 031140 (2012).

\bibitem{Majumder}
S. Majumder and S.K. Das, Phys. Rev. E \textbf{81}, 050102 (2010); ibid \textbf{84}, 021110 (2011).

\bibitem{Khanna}
R. Khanna, N.K. Agnihotri, M. Vashistha, A. Sharma, P.K. Jaiswal and S. Puri, Phys. Rev. E \textbf{82}, 011601 (2010).

\bibitem{Lippiello}
E. Lippiello, A. Mukherjee, S. Puri and M. Zannetti, Europhys. Lett. \textbf{90}, 46006 (2010). F. Corberi, E. Lippiello, A. Mukherjee, S. Puri and M. Zannetti,
J. Stat. Mech., P03016 (2011);   
Phys. Rev. E \textbf{85}, 021141 (2012). 

\bibitem{Hore}
M.J.A. Hore and M. Laradji, J. Chem. Phys. \textbf{132}, 024908 (2010).

\bibitem{Bucior}
K. Bucior, L. Yelash and K. Binder, Phys. Rev. E \textbf{77}, 051602 (2008).

\bibitem{Sicilia}
A. Sicilia, Y. Sarrazin, J.J. Arenzon, A.J. Bray and L.F. Cugliandolo, Phys. Rev. E \textbf{80}, 031121 (2009).

\bibitem{Blond}
N. Blondiaux, S. Morgenthaler, R. Pugin, N. D. Spencer and M. Liley, Appl. Surf. Sci. \textbf{254}, 6820 (2008);
J. Liu, X. Wu, W. N. Lennard and D. Landheer, Phys. Rev. B \textbf{80}, 041403 (2009).

\bibitem{Mitchel}
S.J. Mitchell and D.P. Landau, Phys. Rev. Lett. \textbf{97}, 025701 (2006).

\bibitem{Das1}
S.K. Das, S. Puri, J. Horbach and K. Binder, Phys. Rev. E \textbf{72}, 061603 (2005);
ibid \textbf{73}, 031604 (2006); Phys. Rev. Lett. \textbf{96}, 016107 (2006).

\bibitem{Castellano}
C. Castellano, F. Corberi and M. Zannetti,
Phys. Rev. E  \textbf{56}, 4973 (1997).
F. Corberi and C. Castellano, ibid \textbf {58}, 4658 (1998).
C. Castellano and F. Corberi, ibid \textbf{57}, 672 (1998); ibid \textbf{61}, 3252 (2000);
Phys. Rev. B \textbf{63}, 060102 (2001).

\bibitem{Salernogroup}
F. Corberi, E. Lippiello and M. Zannetti, Phys. Rev. E \textbf{65}, 046136 (2002);
ibid {\bf 78}, 011109 (2008). F. Corberi, A. Coniglio and M. Zannetti, ibid \textbf{51},
5469 (1995).

\bibitem{Puri1}
S. Puri and V. Wadhawan (Eds.), \textit{Kinetics of Phase Transitions} (CRC Press, Boca Raton, FL, 2009).

\bibitem{Bray}
A.J. Bray, Adv. Phys. \textbf{51}, 481 (2002).

\bibitem{Cugliandolo}
J.P. Bouchaud, L.F. Cugliandolo, J. Kurchan and M. Mezard in \textit{Spin Glasses and Random fields, Directions in Condensed Matter Physics} Vol. 12, p. 161, A.P. Young (Ed.) (World Scientific, Singapore, 1998).
F. Corberi, L.F. Cugliandolo, H. Yoshino, in 
\textit{Dynamical heterogeneities in glasses, colloids, and granular media}, Eds.: L. Berthier, G. Biroli, J-P Bouchaud, L. Cipelletti and W. van Saarloos (Oxford University Press, 2011). M. Zannetti, in \cite{Puri1}, p. 153.

\bibitem{Binder1}
K. Binder, in \textit{Phase Transformation of Materials}, Vol. \textbf{5}, p. 405, R.W. Cahn, P. Haasen and E.J. Kramer (Eds.) (VCH, Weinheim, 1991).

\bibitem{Jones}
R.A.L. Jones, \textit{Soft Condensed Matter} (Oxford University Press, Oxford, 2008).
A. Onuki, \textit{Phase Transition Dynamics} (Cambridge University Press, Cambridge, 2002).

\bibitem{LS}
I.M. Lifshitz and V.V. Slyozov, J. Phys. Chem. Solids \textbf{19}, 35 (1961).

\bibitem{Siggia}
E.D. Siggia, Phys. Rev. A \textbf{20}, 595 (1979).
H. Furukawa, Phys. Rev. A \textbf{31}, 1103 (1985); ibid \textbf{36}, 2288 (1987).

\bibitem{Huse}
D.A. Huse, Phys. Rev. B \textbf{40}, 304 (1989).
D.S. Fisher and D.A. Huse, Phys. Rev. B \textbf{38}, 373 (1989).

\bibitem{YRD}
F. Liu, and G.F. Mazenko, Phys. Rev. B {\bf 44}, 9185 (1991);
C. Yeung, M. Rao and R.C. Desai, Phys. Rev. E \textbf{53}, 3073 (1996);
G.F. Mazenko, Phys. Rev. E {\bf 58}, 1543 (1998).

\bibitem{allen}
M.P. Allen and D.J. Tildesley, \textit{Computer Simulations of Liquids}, (Clarendon, Oxford, 1987).

\bibitem{frenkel}
D. Frankel and B. Smit, \textit{Understanding Molecular Simulations: From Algorithms to Application} (Academic Press, San Diego, California, 2002).

\bibitem{Das2}
S.K. Das, M.E. Fisher, J.V. Sengers, J. Horbach and K. Binder, Phys. Rev. Lett. \textbf{97}, 025702 (2006).
S.K. Das, J. Horbach, K. Binder, M.E. Fisher and J. V. Sengers, J. Chem. Phys. \textbf{125}, 024506 (2006).
S. Roy and S.K. Das, Europhys. Lett. \textbf{94}, 36001 (2011).

\bibitem{new}

S.K. Das, S. Roy, S. Majumdar, and S. Ahmad, Europhys. Lett. \textbf{97}, 66006 (2012). 

\bibitem{lara} M. Laradji, S. Toxvaerd and O.G. Mouritsen, Phys. Rev. Lett. \textbf{77}, 2253 (1996).

\bibitem{kab} H. Kabrede and R. Hentschke, Physica A \textbf{361}, 485 (2006).

\bibitem{thak} A.K. Thakre, W.K. den Otter and W.J. Briels, Phys. Rev. E \textbf{77}, 011503 (2008).

\end{thebibliography}
\end{document}